\documentclass[10pt, conference]{IEEEtran}
\usepackage{graphicx, xcolor} 
\usepackage{amsmath,amssymb,epsfig,psfrag,cite}
\usepackage{tikz}
\usepackage{balance}
\usepackage[caption=false,font=footnotesize]{subfig}
\usepackage{pgfplots}
\pgfplotsset{compat=1.18}
\pgfplotsset{
        every axis legend/.append style={
            font=\footnotesize
        }
    }
\newlength\figureheight
\newlength\figurewidth
\usetikzlibrary{arrows.meta, calc, positioning, patterns}

\title{Hybrid Monostatic and Bistatic Sensing for Target Localization in OFDM Systems under Phase Noise}
\author{\IEEEauthorblockN{Luc\'ia Pallar\'es Rodr\'iguez\IEEEauthorrefmark{1},  Musa Furkan Keskin\IEEEauthorrefmark{2}, Alireza Pourafzal\IEEEauthorrefmark{2}, \\ Jos\'e A. L\'opez Salcedo\IEEEauthorrefmark{1}, Gonzalo Seco-Granados\IEEEauthorrefmark{1}, Henk Wymeersch\IEEEauthorrefmark{2}} \vspace{0.1cm}
\IEEEauthorblockA{\IEEEauthorrefmark{1}Dept. of Telecommunications and Systems Engineering, Universitat Aut\`onoma de Barcelona, Spain} 
\IEEEauthorblockA{\IEEEauthorrefmark{2}Dept. of Electrical Engineering, Chalmers University of Technology, Sweden
}\thanks{This work was supported by the Catalan Agency for Management of University and Research Grants (AGAUR) under Grant 2024FI-100186, Societat Econ\`{o}mica Barcelonesa d'Amics del Pa\'is, the SNS JU project 6G-DISAC under the EU's Horizon Europe research and innovation Program under Grant Agreement No 101139130, and the Swedish Research Council (VR) through the project 6G-PERCEF under Grant 2024-04390.}
}

\newcommand{\Tsym}{T_{\text{sym}}}
\newcommand{\Tcp}{T_{\text{CP}}}
\newcommand{\fosc}[1]{f_{\text{3dB}}^{#1}}
\newcommand{\dpn}[1]{\boldsymbol{\xi}_{{#1}}}
\newcommand{\dpntau}[1]{\boldsymbol{\xi}_{{#1}}\!\left(\tau_{{#1}}\right)}
\newcommand{\dpnS}[2]{\xi_{{#1}}\!\left(#2, \tau_{#1}\right)}
\newcommand{\dpnIncS}[2]{\widetilde{\xi}_{{#1}}\!\left(#2\right)}
\newcommand{\dpnInc}{\widetilde{\boldsymbol{\xi}}}
\newcommand{\dpnIncBar}{\bar{\boldsymbol{\xi}}}
\newcommand{\y}[1]{\mathbf{y}_{{#1}}}
\newcommand{\yVect}{\mathbf{y}}
\newcommand{\yPNfree}[1]{\mathbf{y}^{\text{PN-free}}_{{#1}}}
\newcommand{\yPNfreeVect}{\mathbf{y}^{\text{PN-free}}}
\newcommand{\muPNfree}{\boldsymbol{\mu}^{\text{PN-free}}}
\newcommand{\muVect}{\boldsymbol{\mu}}
\newcommand{\p}[1]{\mathbf{p}_{{#1}}}
\newcommand{\pUE}{\mathbf{p}}
\newcommand{\walpha}[1]{\widetilde{\alpha}_{#1}}
\newcommand{\uell}{u_\ell}

\newcommand{\yDFC}{\mathbf{y}}

\newcommand{\x}[1]{\mathbf{x}_{{#1}}}
\newcommand{\Q}{\mathbf{Q}}
\newcommand{\qTau}[1]{\mathbf{q}_{{#1}}(\tau_{{#1}})}
\newcommand{\qP}[1]{\mathbf{q}_{{#1}}(\pUE)}
\newcommand{\eye}[1]{\mathbf{I}_{#1}}
\newcommand{\nVect}{\mathbf{z}}
\newcommand{\nVar}{\sigma^2_{z}}
\newcommand{\pndel}[3]{\phi^{{#1}}\!\left( #2 - \tau_{{#3}}\right)}
\newcommand{\pn}[2]{\phi^{{#1}}\!\left( #2 \right)}
\newcommand{\bstV}[1]{\mathbf{b}\left(\tau_{#1}\right)}
\newcommand{\wtau}[1]{\mathbf{w}_{{#1}}\!\left(\tau_{{#1}}\right)}
\newcommand{\w}[1]{\mathbf{w}_{{#1}}}
\newcommand{\widew}[1]{\widetilde{\mathbf{w}}_{#1}}
\newcommand{\Nsc}{N}
\newcommand{\complex}[1]{\mathbb{C}^{#1}}
\newcommand{\real}[1]{\mathbb{R}^{#1}}
\newcommand{\realPart}[1]{\mathfrak{R}\{#1\}}
\newcommand{\realPartlarge}[1]{\mathfrak{R}\Biggl\{#1\Biggr\}}
\newcommand{\imagPart}[1]{\mathfrak{I}\{#1\}}
\newcommand{\alphaVect}{\boldsymbol{\alpha}}
\newcommand{\fft}{\mathbf{F}}
\newcommand{\fim}{\mathbf{J}}
\newcommand{\hfim}{\fim^{\text{hyb}}}
\newcommand{\hfimO}{\fim^{\text{(o)}}}
\newcommand{\hfimP}{\fim^{\text{(p)}}}
\newcommand{\trans}{\mathsf{T}}
\newcommand{\herm}{\mathsf{H}}
\newcommand{\CN}[2]{\mathcal{CN}\left(#1,#2\right)}
\newcommand{\N}[2]{\mathcal{N}\big(#1,#2\big)}
\newcommand{\expec}[1]{\mathbb{E}\bigl\{#1\bigr\}}
\newcommand{\expecPNlarge}[1]{\mathbb{E}_{p}\left\{#1\right\}}

\newcommand{\expecDPNres}[1]{\mathbb{E}_{p_{\dpnIncBar}}\bigl\{#1\bigr\}}
\newcommand{\expecDPNreslarge}[1]{\mathbb{E}_{p_{\dpnIncBar}}\left\{#1\right\}}
\newcommand{\RdpnInc}{\mathbf{R}_{\dpnIncBar}}
\newcommand{\unk}{\boldsymbol{\eta}}
\newcommand{\unkH}{\boldsymbol{\eta}^{\text{hyb}}}
\newcommand{\crb}{\text{CRB}}
\newcommand{\mcrb}{\text{MCRB}}
\newcommand{\lb}{\text{LB}}
\newcommand{\peb}{\text{PEB}}
\newcommand{\pebhyb}{\text{PEB}^{\text{hyb}}}
\newcommand{\norm}[1]{\lVert #1 \rVert}

\newcommand{\Amcrb}{\mathbf{A}_{\unk_0}}
\newcommand{\Bmcrb}{\mathbf{B}_{\unk_0}}

\DeclareMathOperator*{\argmin}{arg\,min}
\DeclareMathOperator*{\argmax}{arg\,max}

\definecolor{AlirezaPurple}{RGB}{150, 0, 250}

\IEEEoverridecommandlockouts

\begin{document}
\twocolumn[
\begin{@twocolumnfalse}
{\Large{\textbf{IEEE Copyright Notice}}}

\vspace{1em}
\textcopyright 2026 IEEE.  Personal use of this material is permitted.  Permission from IEEE must be obtained for all other uses, in any current or future media, including reprinting/republishing this material for advertising or promotional purposes, creating new collective works, for resale or redistribution to servers or lists, or reuse of any copyrighted component of this work in other works.
\vspace{1em}
\end{@twocolumnfalse}
]
\maketitle
\begin{abstract}
    In hybrid monostatic--bistatic orthogonal frequency-division multiplexing (OFDM) sensing, each sensing node receives a superposition of monostatic and bistatic target returns, whose phase-noise (PN) characteristics differ fundamentally due to oscillator sharing. While monostatic sensing benefits from partial PN cancellation through the common transmit--receive oscillator, bistatic sensing involves independent oscillators and is therefore more strongly affected by PN. In this paper, we characterize the joint residual differential PN statistics of these superimposed components in a centralized two-node sensing system and quantify their impact on target localization via theoretical lower bounds considering three different receivers: (i) PN-free ideal receiver, (ii) PN-aware receiver that exploits the joint PN statistics, and (iii) PN-ignorant receiver that neglects PN. 
The results show that incorporating bistatic observations benefits the PN-aware receiver by providing additional positioning information with negligible PN-induced loss, whereas for the PN-ignorant receiver, the detrimental effect of the large unmodeled PN variance associated with bistatic reflections outweighs this additional information and degrades estimation accuracy.
    
\end{abstract}


\section{Introduction}\label{sec:intro}

Sensing is expected to play a crucial role in next-generation wireless networks, particularly through the integrated sensing and communications (ISAC) paradigm \cite{Liu_22}. In dual-functional ISAC systems, communication waveforms and hardware are reused to perform sensing and extract environmental information, enabling efficient use of spectrum and infrastructure \cite{Zhang_22}. This trend is particularly relevant at mmWave frequencies, where large bandwidths provide fine delay resolution. However, operation at high carrier frequencies makes the system more sensitive to hardware impairments. In this context, oscillator phase noise (PN) can become a major bottleneck, especially in orthogonal frequency-division multiplexing (OFDM) systems, which are promising candidates for ISAC \cite{Bica_16}.

The impact of PN on OFDM sensing has recently been studied for monostatic and bistatic configurations
\cite{PN_OFDM_ISAC_RadarConf_2023, OFDM_PN_TSP_Exploitation_2023, han_25, oliveira_24, Idrees_26}.
In particular, \cite{han_25} considers a distributed OFDM ISAC system and investigates the effects of synchronization errors and PN under centralized, decentralized, and monostatic-only processing. However, the PN processes associated with the different sensing links are not treated jointly as random nuisance parameters in the localization problem.

In this work, we explicitly characterize the joint residual differential phase noise (DPN) process across the monostatic and bistatic observations. The resulting covariance captures the distinct PN behavior of monostatic and bistatic links, as well as the statistical coupling between links that share the same oscillators. To quantify the localization impact of this joint DPN model in the centralized processing scheme, we consider two receiver assumptions. For a PN-aware receiver, the residual DPN is incorporated as a random nuisance parameter and the hybrid Cram\'er--Rao bound (HCRB) is derived. For a PN-ignorant receiver, which processes the observations using a PN-free likelihood, we derive the misspecified CRB (MCRB) that captures both the estimation variance and the bias induced by the model mismatch. A PN-free CRB is additionally used as an ideal benchmark. Finally, by comparing hybrid monostatic--bistatic processing with monostatic-only processing, we quantify when processing the bistatic components jointly with the monostatic returns improves localization accuracy and how this depends on whether the receiver accounts for PN.


\vspace{-0.15in}
\section{Hybrid Sensing System Model}\label{sec:sys_model}

\subsection{Scenario Description}

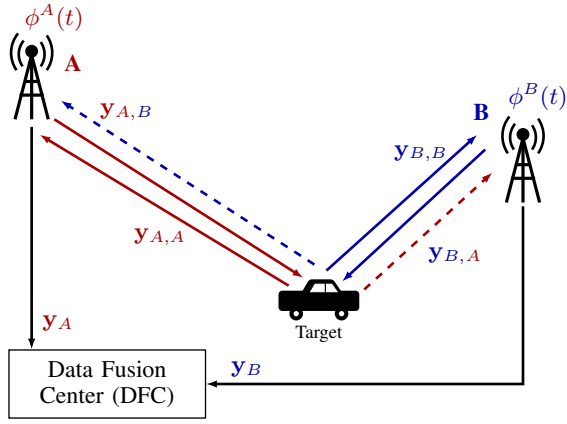
\begin{figure}[t!]
    \centering
    \begin{tikzpicture}[
    scale=1.0,
    transform shape,
    >=Latex,
    every node/.style={font=\rmfamily},
    arrow/.style={-{Latex[length=1.5mm,width=1mm]}, line width=1pt},
    redarrow/.style={arrow, draw=red!65!black},
    bluearrow/.style={arrow, draw=blue!65!black},
    reddashed/.style={redarrow, dashed},
    bluedashed/.style={bluearrow, dashed},
    labelred/.style={text=red!65!black, font=\rmfamily\bfseries\small},
    labelblue/.style={text=blue!65!black, font=\rmfamily\bfseries\small},
    blackarrow/.style={arrow, draw=black},
]

\coordinate (A)   at (-4.0,4.3);
\coordinate (B)   at ( 2.5,3.2);
\coordinate (CAR) at (-0.2,1.5);
\coordinate (DFC) at (-3.0,0.4);

\tikzset{
    basestation/.pic={
        \draw[black, line width=1.2pt] (0,-0.4) -- (0,0.5);
        \draw[black, line width=1.2pt] (-0.22,-0.4) -- (0,0.45) -- (0.22,-0.4);
        \draw[black, line width=1.2pt] (-0.16,-0.08) -- (0.16,-0.08);
        \draw[black, line width=1.2pt] (-0.10,0.10) -- (0.10,0.10);
        \fill[black] (0,0.5) circle (0.085);
        \draw[black, line width=1pt] (0.12,0.30) arc[start angle=-35,end angle=35,radius=0.28];
        \draw[black, line width=1pt] (-0.12,0.30) arc[start angle=215,end angle=140,radius=0.28];
        \draw[black, line width=1pt] (0.20,0.25) arc[start angle=-35,end angle=35,radius=0.4];
        \draw[black, line width=1pt] (-0.20,0.25) arc[start angle=215,end angle=140,radius=0.4];
    }
}

\pic at (A) {basestation};
\pic at (B) {basestation};

\node[labelred]  at ($(A)+(+0.55,0.35)$) {A};
\node[labelred]  at ($(A)+(0.3,0.95)$) {$\phi^A(t)$};

\node[labelblue] at ($(B)+(-0.55,0.8)$) {B};
\node[labelblue] at ($(B)+(0.2,1.0)$) {$\phi^B(t)$};

\tikzset{
    car/.pic={
        \draw[black, fill=black, line width=0.8pt, rounded corners=1pt]
            (-0.52,-0.12) rectangle (0.52,0.12);

        \draw[black, fill=black, line width=0.8pt, rounded corners=0.8pt]
            (-0.23,0.08) -- (-0.12,0.27) -- (0.18,0.27)
            -- (0.31,0.08) -- cycle;

        \draw[white, fill=white, line width=0.3pt]
            (-0.11,0.12) -- (-0.05,0.22) -- (0.07,0.22)
            -- (0.07,0.12) -- cycle;

        \draw[white, fill=white, line width=0.3pt]
            (0.10,0.12) -- (0.10,0.22) -- (0.17,0.22)
            -- (0.24,0.12) -- cycle;

        \fill[black] (-0.3,-0.17) circle (0.09);
        \fill[black] ( 0.3,-0.17) circle (0.09);
        \fill[white] (-0.3,-0.17) circle (0.035);
        \fill[white] ( 0.3,-0.17) circle (0.035);
    }
}

\pic at (CAR) {car};
\node[font=\rmfamily\scriptsize, text=black, below=2mm] at (CAR) {Target};

\node[
    fill=white,
    draw=black,
    line width=0.5pt,
    text=black,
    align=center,
    font=\rmfamily\small,
    minimum width=26mm,
    minimum height=9mm
] (dfc) at (DFC) {Data Fusion\\Center (DFC)};

\coordinate (tAup)   at ($(CAR)+(-0.20, 0.30)$);
\coordinate (tAdown) at ($(CAR)+(-0.40,0.2)$);
\coordinate (tBup)   at ($(CAR)+( 0.1, 0.4)$);
\coordinate (tBdown) at ($(CAR)+( 0.3, 0.25)$);

\draw[redarrow]  ($(A)+(0.3,-0.4)$) -- (tAup);
\draw[redarrow]  (tAdown) -- ($(A)+(0.1,-0.6)$);

\draw[reddashed] ($(CAR)+(0.6,0.15)$) -- ($(B)+(-0.4,0.0)$);

\draw[bluearrow] ($(B)+(-0.5,0.3)$) -- (tBdown);
\draw[bluearrow] (tBup) -- ($(B)+(-0.6,0.5)$);

\draw[bluedashed] ($(CAR)+(-0.06,0.48)$) -- ($(A)+(0.38,-0.15)$);

\node[labelred]  at (-2.3,2.35) {$\mathbf{y}_{A,A}$};
\node[labelred]  at (-2.85, 4) {$\mathbf{y}_{A,}$};
\node[labelblue] at (-2.5, 4) {$_B$};

\node[labelblue] at (1.15,3.45) {$\mathbf{y}_{B,B}$};
\node[labelblue] at (1.50, 2.1) {$\mathbf{y}_{B,}$};
\node[labelred]  at (1.85, 2.1) {$_A$};


\node[labelblue] at ($(DFC)+(1.85,0.2)$) {$\mathbf{y}_B$};
\node[labelred]  at ($(DFC)+(-0.65,0.8)$) {$\mathbf{y}_A$};

\draw[blackarrow] ($(B)+(0,-0.45)$) -- (2.5, 0.4) -- ($(DFC)+(+1.3, 0)$);
\draw[blackarrow] ($(A)+(0,-0.5)$) -- ($(DFC)+(-1, 0.45)$);

\end{tikzpicture}
    \caption{System model for the centralized hybrid sensing scenario, where each node (A, B) captures both the monostatic and bistatic returns in the presence of phase noise ($\phi^A(t), \phi^B(t)$).}
    \label{fig:proposed_scenario}
\end{figure}

We consider the centralized hybrid monostatic-bistatic sensing setup shown in Fig.~\ref{fig:proposed_scenario}, with two sensing nodes, A and B, and a single target. The nodes are connected to a DFC, which provides time and frequency synchronization, and collects the received observations for centralized parameter estimation. Each node transmits an OFDM sensing signal towards the target and receives the superposition of the monostatic echo of its own transmitted signal and the bistatic echo of the signal transmitted by the other node \cite{han_25}. A direct line-of-sight (LoS) link between the nodes is not considered. The transmitter and receiver chains within each node share the same local oscillator (LO), which is independent of the LO at the other node \cite{PN_OFDM_ISAC_RadarConf_2023,OFDM_PN_TSP_Exploitation_2023, han_25}. Due to hardware imperfections, all the LOs are considered to be imperfect and impaired by PN \cite{Petrovic_07}. 

Let $\p{A} \in \mathbb{R}^2$ and $\p{B} \in \mathbb{R}^2$ denote the known locations of nodes A and B, respectively, and let $\pUE \in \mathbb{R}^2$ denote the unknown target position. The monostatic and bistatic propagation delays are given by
\begin{equation}
\begin{aligned}
      \tau_A = 2\norm{\p{A} - \pUE}/{c},   \ \tau_B = 2{\norm{\p{B} - \pUE}}/{c},    \\
\tau_{A,B} = \tau_{B,A} = {\norm{\p{A} - \pUE}}/{c} + {\norm{\p{B} - \pUE}}/{c},    
\end{aligned}    
\label{eq:geom_rel}
\end{equation}
where $c$ is the speed of light.


\subsection{Transmit and Received Signal Models}

Let $p, q\in \{A, B\}$ denote the receiving and transmitting nodes, respectively. The nodes transmit simultaneously, each transmitting an OFDM symbol with $\Nsc$ subcarriers. The total symbol duration is $\Tsym=\Tcp + T$, where $\Tcp$ and $T$ are the cyclic prefix (CP) and useful symbol duration, respectively. The subcarrier spacing is denoted by $\Delta f = 1/T$, leading to a total bandwidth of $B=\Nsc\Delta f$. The baseband OFDM signal transmitted by node $q$ can be expressed as \cite{Bica_16} $s_q(t)=\frac{1}{\sqrt{\Nsc}}\sum_{n=0}^{\Nsc-1}(x_n)_q\,  e^{j2\pi n\Delta f t}\, \Pi\left(\frac{t}{\Tsym}\right)$, where $x_n$ is the complex data on the $n$-th subcarrier and $\Pi\left(t\right)$ is a rectangular pulse with $\Pi\left(t\right) = 1$ for $t\in [0, 1]$ and zero otherwise. In the presence of PN, the signal transmitted by node $q$ is upconverted through $e^{j\left[2\pi f_c t + \pn{q}{t} \right]}$, where $f_c$ is the carrier frequency and $\pn{q}{t}$ is the PN process of its LO. After propagation and downconversion at node $p$ with $e^{-j\left[2\pi f_c t + \pn{p}{t}\right]}$, the received baseband signal is \cite{PN_OFDM_ISAC_RadarConf_2023,OFDM_PN_TSP_Exploitation_2023}
\begin{equation}
        y_{p, q}(t) =\alpha_{p, q}s_{q}(t - \tau_{p, q})\, w_{p, q}(t,\tau_{p, q}),
    \label{eq:RxBBsig_pq}
\end{equation}
where $\alpha_{p,q}$ and $\tau_{p, q}$ denote the complex channel gain and propagation delay associated with the sensing link from transmitter $q$ to receiver $p$, respectively, and $w_{p, q}(t, \tau_{p, q}) = e^{-j\, \dpnS{p, q}{t}}$ is the multiplicative PN term, with
\begin{equation}
   \dpnS{p, q}{t} = \pn{p}{t} - \pndel{q}{t}{p,q},
  \label{eq:dpn_def}
\end{equation}
denoting the DPN associated with the sensing link from node $q$ to node $p$. Sampling \eqref{eq:RxBBsig_pq} at $t = \Tcp + \ell\, T / \Nsc$ for $\ell = 0, \ldots, \Nsc-1$, i.e., after CP removal, and assuming $\Tcp \geq \tau_{p,q}$ \cite{Firat_OFDM_2012}, the discrete-time observation vector is 
\begin{equation}
    \y{p, q} =\alpha_{p,q} \wtau{p, q} \odot \fft^{\herm}\left(\x{q} \odot \bstV{p, q}\right)\ \in \complex{\Nsc\times 1},
    \label{eq:RxBBsig_pq_vect}
\end{equation}
where $[\y{p, q} ]_\ell \triangleq y_{p, q}(\Tcp + \ell\, T / \Nsc)$, $\x{q} = \left[(x_0)_q, \ldots , (x_{N-1})_q\right]^\trans \in \complex{\Nsc\times 1}$, $\fft \in \complex{\Nsc\times \Nsc}$ is the unitary DFT matrix, whose $(\ell, n)$-th entry is $\left[ \fft\right]_{\ell,n} = \frac{1}{\sqrt{\Nsc}} e^{- j 2 \pi n \frac{\ell}{\Nsc}} $, and $\odot$ denotes the Hadamard product. Moreover, $\wtau{p,q} \in \complex{\Nsc\times 1}$ is the multiplicative PN vector, with $[\wtau{p,q}]_\ell \triangleq  w_{p, q}(t, \tau_{p, q})\rvert_{t = \Tcp + \ell\, T / \Nsc}$, and $\bstV{p, q} \triangleq   \left[ 1, e^{-j 2 \pi \Delta f \tau_{p, q}}, \ldots,  e^{-j 2 \pi (\Nsc-1) \Delta f  \tau_{p, q}} \right]^\trans$ is the frequency-domain steering vector associated with the sensing link from node $q$ to node $p$.

\subsection{Hybrid Sensing Observations}

The generic signal model in \eqref{eq:RxBBsig_pq_vect} can be particularized to the monostatic and bistatic sensing links by properly selecting the node indices ($p$, $q$). For notational simplicity, we drop the explicit dependence of the multiplicative PN vector and DPN vector on the propagation delay and write $\w{p, q} \triangleq \wtau{p, q}$ and $\dpn{p, q} \triangleq \dpntau{p, q}$, respectively. For monostatic links, i.e., $p=q$, we use the simplified notation $\alpha_{p}\triangleq\alpha_{p, p}$, $\w{p} \triangleq\w{p, p}$, and $\tau_{p} \triangleq\tau_{p,p}$.
Since both nodes operate within the same time-frequency resources, each node receives the superposition of its target-reflected monostatic component and the bistatic component from the other node. Therefore, the overall received signal at node A is given by
\begin{align}
    \y{A} & = \y{A, A} + \y{A, B} + \nVect_A  \label{eq:recSig_A}\\ & = \alpha_{A} \w{A} \odot \qTau{A} + \alpha_{A, B} \w{A, B} \odot \qTau{A, B} + \nVect_A,\notag     
\end{align}
where $\qTau{p} = \fft^{\herm}\left(\x{p} \odot \bstV{p}\right)$, $\qTau{p, q} = \fft^{\herm}\left(\x{q} \odot \bstV{p, q}\right)$, and $\nVect_A \sim \CN{0}{\nVar\eye{\Nsc}}$. Equivalently, the received signal at node B is given by
\begin{align}
    \y{B} & = \y{B, B} + \y{B, A} + \nVect_B \label{eq:recSig_B}\\ & = \alpha_{B} \w{B} \odot \qTau{B} + \alpha_{B, A} \w{B, A} \odot \qTau{B, A} + \nVect_B,\notag    
\end{align}
where $\nVect_B \sim \CN{0}{\nVar\eye{\Nsc}}$. The hybrid observations in \eqref{eq:recSig_A} and \eqref{eq:recSig_B} are forwarded to the DFC and stacked into a centralized observation vector $\yDFC = \left[\y{A}^\trans\ \y{B}^\trans \right]^\trans \in \complex{2\Nsc \times 1}$.

\section{Phase Noise Statistics}\label{sec:phase_noise}

This section provides a statistical characterization of the PN processes at the nodes, which will be used in the derivation of the performance bounds in the following section. Under the free-running oscillator model considered in this work, the PN $\pn{j}{t}$ at node $j$, with $j\in \{A, B\}$, is modeled as a Wiener process, so that for a given time instant $t$ \cite{OFDM_PN_TSP_Exploitation_2023,Petrovic_07}, $\pn{j}{t} \sim \N{0}{4\pi \fosc{j}\, t},$
where $\fosc{j}$ is the 3-dB bandwidth of the Lorentzian oscillator spectrum at node $j$, and the autocorrelation function is given by $\expec{\pn{j}{t}\pn{j}{t + \tau}}=4\pi \fosc{j}\min(t, t + \tau)$ \cite{Petrovic_07}. 

\subsection{Reparametrization of the Signal Model}

Based on the DPN definition in \eqref{eq:dpn_def} and the sampling instants, we decompose the monostatic and bistatic DPN processes at $t_\ell = t_0 + \uell$, where $t_0 = \Tcp$ and $\uell = \ell\, T/\Nsc \geq 0 $, into their initial value at $t_0$ and a residual increment
\begin{subequations}
\begin{align}
   \dpnS{p}{t_0 \!+\! \uell} \! & = \! \underbrace{\pn{p}{t_0} \!-\! \pndel{p}{t_0}{p}}_{\text{Initial phase offset}} + \dpnIncS{p}{\uell, \tau_p},
   \label{eq:dpn_rep_mono}
    \\
   \dpnS{p, q}{t_0 \!+\! \uell} \!  & = \!  \underbrace{\pn{p}{t_0} \!-\! \pndel{q}{t_0}{p, q}}_{\text{Initial phase offset}} + \dpnIncS{p, q}{\uell, \tau_{p, q}},
   \label{eq:dpn_rep_bi}
\end{align}
\end{subequations}
where
\begin{subequations}
\begin{multline}
   \dpnIncS{p}{\uell,\tau_p} = \left[\pn{p}{t_0 + \uell} \!-\! \pn{p}{t_0}\right] \\ - \left[\pndel{p}{t_0 + \uell}{p} \!-\! \pndel{p}{t_0}{p}\right],
   \label{eq:dpn_res_mono}
\end{multline}
\vspace{-0.8cm}
\begin{multline}
   \dpnIncS{p, q}{\uell, \tau_{p, q}} = \left[\pn{p}{t_0 + \uell} \!-\! \pn{p}{t_0}\right] \\ - \left[\pndel{q}{t_0 + \uell}{p, q} \!-\! \pndel{q}{t_0}{p,q}\right].
   \label{eq:dpn_res_bi}
   \end{multline}
   \label{eq:dpn_res}
\end{subequations}

The initial phase offsets in \eqref{eq:dpn_rep_mono} and \eqref{eq:dpn_rep_bi} are independent of the time offset $\uell$. Hence, they induce a constant phase rotation across the $\Nsc$ samples of the monostatic and bistatic observations. These rotations can be absorbed into the unknown channel gains $\alpha_{p}$ and $\alpha_{p, q}$, respectively. As a result, the monostatic and bistatic observations are reparametrized as
\begin{subequations}
\begin{align}
    \y{p, p} & =\walpha{p} \widew{p} \odot \qTau{p},
    \label{eq:sig_mono_reparam}
    \\
    \y{p, q} &=\walpha{p, q} \widew{p, q} \odot \qTau{p, q},
    \label{eq:sig_bi_reparam}
\end{align}
\end{subequations}
where
\begin{subequations}
\begin{align}
    \walpha{p} &\triangleq \alpha_{p}\, e^{-j\left[\pn{p}{t_0} - \pndel{p}{t_0}{p}\right]}, \  \left[\widew{p}\right]_\ell \triangleq e^{-j\dpnIncS{p}{\uell, \tau_p} },
    \\
    \walpha{p, q}&\triangleq \alpha_{p, q}\, e^{-j\left[\pn{p}{t_0} - \pndel{q}{t_0}{p, q}\right]}, \ \left[\widew{p, q}\right]_\ell  \triangleq  e^{-j\dpnIncS{p, q}{\uell,  \tau_{p, q}} }.
\end{align}
\end{subequations}


We define $\dpnInc_p$, $\dpnInc_{p,q} \in\real{\Nsc\times 1}$ such that $\widew{p} = e^{-j\dpnInc_p}$, $\widew{p,q} =e^{-j\dpnInc_{p,q}}$, and stack these contributions into the residual DPN vector, $\dpnInc = \bigl[\dpnInc_A^\trans\ \dpnInc_{A,B}^\trans\ \dpnInc_{B,A}^\trans\ \dpnInc_B^\trans\bigr]^\trans \in \real{4\Nsc\times 1}$. The dependence of $\dpnInc$ on the propagation delays is kept implicit.



\subsection{Statistics of the Residual DPN}\label{subsec:PN-stat}

To characterize the joint statistics of the residual DPN vector $\dpnInc$, let us first write the residual DPN processes as
\begin{subequations}    
\begin{align}
    \left[\dpnInc_A\right]_{\ell} & = \Delta_A(0, \uell) - \Delta_A(\tau_A, \uell),\\
    \left[\dpnInc_{A, B}\right]_{\ell} & = \Delta_A(0, \uell) - \Delta_B(\tau_{A,B}, \uell),\\
    \left[\dpnInc_{B, A}\right]_{\ell} & = \Delta_B(0, \uell) - \Delta_A(\tau_{B,A}, \uell),\\
    \left[\dpnInc_B\right]_{\ell} & = \Delta_B(0, \uell) - \Delta_B(\tau_B, \uell),
    \label{eq:def_resDPN_delta}
\end{align}
\end{subequations}
where $ \Delta_j(\tau, \uell) = \pn{j}{t_0 + \uell - \tau} - \pn{j}{t_0 - \tau},\ j\in\{A, B\},$
denotes the PN increment over the interval $[t_0 - \tau, t_0 - \tau + \uell]$ with self-covariance
\begin{multline}
\expec{\Delta_j(a, u_m) \Delta_j(b, u_n) }\! =\! 4\pi\fosc{j}\!\left[\min(u_m\! -\! a, u_n\! -\!b ) \right. \\\left. \!-\! \min(u_m\!-\!a, -b)\! -\! \min(-a, u_n \!-\! b )\! +\! \min(-a, -b)\right],
\label{eq:cov_Delta}
\end{multline}
for $a, b \in \mathbb{R}$ and $\expec{\Delta_j(a, u_m) \Delta_i(b, u_n) }\!=\! 0$ for $j\!\neq\! i$, since increments from different oscillators are independent.

Applying \eqref{eq:cov_Delta} to the residual DPN processes reveals that the marginal variances for the monostatic and bistatic links are given by $\operatorname{var}\!\big(\dpnIncS{p}{\uell,\tau_p}\big) = 8\pi\fosc{p}\min(\uell, \tau_p)$ and $\operatorname{var}\!\big({\dpnIncS{p, q}{\uell,\tau_{p, q}}}\big) = 4\pi\big(\fosc{p} + \fosc{q}\big)\uell$, respectively. Therefore, monostatic residual PN is delay-dependent and saturates at a value proportional to the propagation delay due to the use of the same LO for upconversion and downconversion \cite{PN_OFDM_ISAC_RadarConf_2023}. In contrast, bistatic residual PN has no such common-oscillator cancellation because the transmitter and receiver LOs are independent, thus its variance grows linearly with $u_\ell$ and is independent of the propagation delay.

Note from \eqref{eq:dpn_res} that, since $u_0 = 0$, $[\dpnInc_p]_0 = [\dpnInc_{p,q}]_0 = 0$ for all sensing links. Hence, the first entry of each of the four residual-DPN vectors is deterministic, and
the covariance matrix of the full vector $\dpnInc$ is singular. To avoid this, we define the reduced residual DPN vector $\dpnIncBar \in \real{4(\Nsc - 1) \times 1}$ by removing the deterministic $\ell = 0$ entry from each of the four residual DPN vectors. Its covariance matrix is obtained from $\eqref{eq:cov_Delta}$ as $\RdpnInc(\boldsymbol{\tau})\! =\! \expec{\dpnIncBar\,  \dpnIncBar^\trans}$. Using the geometric relations in \eqref{eq:geom_rel}, this dependence can be equivalently written in terms of the target position, i.e., $\RdpnInc(\pUE) \triangleq \RdpnInc(\boldsymbol{\tau}(\pUE))\in \real{4(\Nsc - 1)\times 4(\Nsc - 1)}$\footnote{The covariance matrix $\RdpnInc(\pUE)$ is assumed non-singular for the considered geometries and oscillator parameters.}. Therefore, $\dpnIncBar$ can be statistically characterized as
\begin{equation}
   \dpnIncBar \sim \N{\mathbf{0}}{\RdpnInc(\pUE)}.
   \label{eq:dpnInc_dist}
\end{equation} 


\section{CRB Analysis under Phase Noise}\label{sec:bounds}

This section derives the deterministic CRB in the absence of PN as a benchmark, followed by the MCRB and HCRB under PN for estimators that neglect and exploit the PN statistics, respectively. In the following, delay-dependent quantities are expressed in the location domain through \eqref{eq:geom_rel}. In particular,
\begin{subequations}
\begin{align}
    \qP{p}
    &\triangleq
    \left.\qTau{p}\right|_{\tau_p=\tau_p(\pUE)}, \\ 
    \qP{p,q}
    &\triangleq
    \left.\qTau{p,q}\right|_{\tau_{p,q}=\tau_{p,q}(\pUE)}. 
\end{align}
\end{subequations}

\subsection{Deterministic CRB without Phase Noise}

For perfect oscillators, i.e., $\fosc{A} = \fosc{B} = 0$, the received signals at the nodes are PN-free and are given by
\begin{subequations} \label{eq_pn_free_y}
\begin{align}
    \yPNfree{A} & = \alpha_{A}\, \qP{A} + \alpha_{A, B}  \, \qP{A, B} + \nVect_A,
    \label{eq:recSig_PNfree_A}
    \\
    \yPNfree{B} &= \alpha_{B}\, \qP{B} + \alpha_{B, A}  \, \qP{B, A} + \nVect_B.
    \label{eq:recSig_PNfree_B}
\end{align}
\end{subequations}

The two observations are stacked at the DFC as
\begin{equation}
    \yPNfreeVect = \bigl[\left(\yPNfree{A}\right)^\trans \left(\yPNfree{B}\right)^\trans \bigr]^\trans \in \complex{2\Nsc\times 1}.
    \label{eq:recSig_DFC_PNfree}
\end{equation}

The unknown deterministic parameter vector associated with this signal model is defined as
$\unk = \left[\pUE^\trans\ \realPart{\alphaVect^\trans}\ \imagPart{\alphaVect^\trans}\right]^\trans \in \real{10\times 1},$
where $\alphaVect = \left[\alpha_{A}\ \alpha_{A, B}\ \alpha_{B, A}\ \alpha_{B}\right]^\trans$ is the complex channel gain vector, and $\realPart{\alphaVect}$ and $\imagPart{\alphaVect}$ denote its real and imaginary parts, respectively. For any unbiased estimator $\widehat{\unk}(\yPNfreeVect)$ of $\unk$, its covariance matrix is lower bounded by the CRB $\fim^{-1}$, where $\fim\in\real{10\times10}$ is obtained via the Slepian--Bangs formula
\cite[Eq. (15.52)]{kay} as
\begin{equation}
    \bigl[\fim\bigr]_{i, k} = \frac{2}{\nVar} \mathfrak{R}\left\{\frac{\partial \bigl(\muPNfree(\unk)\bigr)^\herm}{\partial \eta_i}\, \frac{\partial \muPNfree(\unk)}{\partial \eta_k} \right\},
    \label{eq:FIM_PNfree}
\end{equation}
where $\eta_n = \bigl[\unk\bigr]_n$ and $\muPNfree(\unk)$ is the noise-free mean vector of \eqref{eq:recSig_DFC_PNfree}. Using \eqref{eq:FIM_PNfree}, the position error bound (PEB) can be derived from the CRB on position estimation as $\peb = \sqrt{\operatorname{tr}\bigl(\crb(\pUE)\bigr)},$
where $\operatorname{tr}(\cdot)$ denotes the trace operator, and $\crb(\pUE)=\bigl[\fim^{-1}\bigr]_{1:2, 1:2}$ is the $2\times 2$ submatrix of $\fim^{-1}$ corresponding to the position parameters.

\subsection{Hybrid CRB under Phase Noise}

The HCRB characterizes the estimation performance when PN is present and the estimator is aware of it. In this case, the estimator uses the PN-impaired centralized observation $\yDFC$, defined after \eqref{eq:recSig_B}, and the hybrid unknown parameter vector is given by $ \unkH=[\pUE^\trans\ \realPart{\alphaVect^\trans}\ \imagPart{\alphaVect^\trans}\ \dpnIncBar^\trans ]^\trans \in \real{\left(4(\Nsc - 1)+10\right)\times 1},$
with both deterministic, $\pUE$ and $\alphaVect$, and random, $\dpnIncBar$, parameters.

The HCRB of an unbiased estimator $\hat{\unk}^{\text{hyb}}(\yDFC)$ of $\unkH$ can be obtained from the inverse of the hybrid FIM (HFIM), defined as $\hfim = \hfimO + \hfimP\ \in \real{\left(4(\Nsc - 1)+10\right)\times \left(4(\Nsc - 1)+10\right)}$, where $\hfimO$ and $\hfimP$ are the observation-related and the prior information-related HFIMs, respectively. Using \cite[Eq. (7)]{Bay_08} and the Slepian-Bangs formula \cite[Eq. (15.52)]{kay}, the $(i, k)$-th entry of $\hfimO$ and $\hfimP$ is given by
\begin{align}
    \left[\hfimO\right]_{i, k} & = \expecDPNreslarge{\frac{2}{\nVar}\realPartlarge{\frac{\partial\muVect^\herm(\unkH)}{\partial\eta^{\text{hyb}}_i}\frac{\partial\muVect(\unkH)}{\partial\eta^{\text{hyb}}_k}}},
    \label{eq:hfimO} \\
    \left[\hfimP\right]_{i, k} & = \expecDPNreslarge{\frac{\partial \log\, p_{\dpnIncBar}(\dpnIncBar; \pUE) }{\partial\eta^{\text{hyb}}_i}\frac{\partial \log\, p_{\dpnIncBar}(\dpnIncBar; \pUE)}{\partial\eta^{\text{hyb}}_k}},
    \label{eq:hfimP}
\end{align}
where $\muVect(\unkH)$ is the mean of the PN-impaired centralized observation $\yDFC$ given $\dpnIncBar$, and $p_{\dpnIncBar}(\dpnIncBar; \pUE)$ is the pdf of the residual DPN as a function of $\pUE$, obtained as
\begin{equation}
    p_{\dpnIncBar}(\dpnIncBar; \pUE)\! =\! \frac{1}{\sqrt{(2\pi)^{4(\Nsc - 1)}\det\bigl(\RdpnInc(\pUE)\bigr)}} \exp{\left\{-\frac{\dpnIncBar^\trans \!\RdpnInc^{-1}\!(\pUE) \dpnIncBar}{2}\right\}}.
    \label{eq:pdf_dpnInc}
\end{equation}

Using \eqref{eq:hfimO} and \eqref{eq:hfimP}, the hybrid PEB is given by $\pebhyb= \sqrt{\operatorname{tr}\Bigl(\Bigl[\bigl(\hfim\bigr)^{-1}\Bigr]_{1:2, 1:2}\Bigr)}.$

\subsection{Misspecified CRB under Phase Noise}

This subsection analyzes the expected performance when PN is present but the receiver is unaware of it and applies processing algorithms designed for PN-free scenarios. To quantify the resulting loss in position estimation accuracy, we use the MCRB \cite{Richmond_15, Fortunati_17}, which characterizes the degradation caused by the mismatch between the true PN-impaired signal model and the PN-free model assumed at the receiver.

\subsubsection{True and Assumed Models} The true model corresponds to PN-impaired observation $\yDFC$. For a given residual DPN realization $\dpnIncBar$, its conditional pdf is
\begin{equation}
    p(\yVect\vert \dpnIncBar) =\frac{1}{(\pi\nVar)^{2\Nsc}}\exp{\left\{ -\frac{\norm{\yVect- \muVect}^2}{\nVar}\right\}},
    \label{eq:pdf_PN}
\end{equation}
where $\muVect$ is the mean vector of the PN-impaired centralized observation $\yDFC$. In practice, PN is typically ignored by the receiver, which evaluates the same observation using the misspecified parametric PN-free likelihood given by \cite{Fortunati_17}
\begin{equation}
    \tilde{p}(\yVect ; \unk) =\frac{1}{(\pi\nVar)^{2\Nsc}}\exp{\left\{ -\frac{\norm{\yVect- \muPNfree(\unk)}^2}{\nVar}\right\}},   
    \label{eq:pdf_PN-free}
\end{equation}
where $\muPNfree(\unk)$ is the mean vector of the assumed PN-free observation model in \eqref{eq:recSig_DFC_PNfree} for a given $\unk$.


\subsubsection{Pseudo-True Parameter}

To derive the MCRB, we first define the pseudo-true parameter vector $\unk_0$ as the minimizer of the Kullback--Leibler divergence (KLD) between the true and assumed models \cite{Fortunati_17}. Based on \cite[Lemma 1]{Ozturk_23}, it is equivalently obtained as
\begin{equation}
    \unk_0 = \argmin_{\unk}\ \norm{\muVect - \muPNfree(\unk)}^2.
    \label{eq:pseudo_true_param}
\end{equation}


From \eqref{eq:pseudo_true_param}, $\alphaVect_0$ can be obtained in closed form as\footnote{We assume that the transmitted sequences and target geometry yield a full-column-rank sensing matrix $\Q(\pUE)$, so that $\Q^\herm\!\left(\pUE\right)\Q\!\left(\pUE\right)$ is non-singular.} 
\begin{equation}
    \alphaVect_0 = \left(\Q^\herm\!\left(\pUE\right)\Q\!\left(\mathbf{p}\right)\right)^{-1} \Q^\herm\! \left(\pUE\right)\muVect = \Q^{\dagger}\!\left(\pUE\right)\muVect, 
    \label{eq:pseudo_true_alpha}
\end{equation}
as function of $\pUE$, where, from \eqref{eq_pn_free_y}, 
\begin{equation}
\mathbf{Q}\!\left(\pUE\right) = \begin{bmatrix}
\qP{A} & \qP{A, B} & \mathbf{0} & \mathbf{0} \\
\mathbf{0} & \mathbf{0} & \qP{B, A} & \qP{B}
\end{bmatrix},
\end{equation}
and $\Q^{\dagger}\!\left(\pUE\right)$ is its Moore-Penrose pseudoinverse. Plugging \eqref{eq:pseudo_true_alpha} into \eqref{eq:pseudo_true_param}, the pseudo-true position parameter $\pUE_0$ can be obtained by solving
\begin{equation}
    \pUE_0 = \argmax_{\pUE}\ \muVect^\herm \Q \!\left(\pUE\right)\Q^{\dagger}\!\left(\pUE\right) \muVect.
    \label{eq:pseudo_true_pos}
\end{equation}

\subsubsection{MCRB and LB Derivation}

Using the pseudo-true parameter vector in \eqref{eq:pseudo_true_param}, the MCRB matrix is given by \cite{Fortunati_17}
\begin{equation}
    \mcrb\bigl(\unk_0\bigr) = \Amcrb^{-1}\Bmcrb\Amcrb^{-1} \in \real{10\times 10},
\end{equation}
where $\Amcrb$ and $\Bmcrb$ are defined as
\begin{align}
    \bigl[\Amcrb\bigr]_{i, k} & = \expecPNlarge{\frac{\partial^2}{\partial\eta_i \partial\eta_k}\, \log\, \tilde{p}(\yVect ; \unk)\Big\rvert_{\unk = \unk_0}}, \\ 
    \bigl[\Bmcrb\bigr]_{i, k} & = \begin{aligned}[t]
    \mathbb{E}_{p}\Bigg\{\frac{\partial}{\partial\eta_i }\, \log\, \tilde{p}(\yVect ; \unk)\,   \frac{\partial}{\partial\eta_k}\, \log\, \tilde{p}(\yVect & ; \unk) \Big\rvert_{\unk = \unk_0} \Bigg\},
    \end{aligned} 
\end{align}
with $\mathbb{E}_{p}\{\cdot\}$ denoting the expectation over the true pdf in \eqref{eq:pdf_PN}. For any misspecified-unbiased estimator $\widehat{\unk}(\yVect)$ of $\unk_0$, the mean-square error matrix with respect to the true parameter vector is lower bounded by $\lb(\unk_0)$, defined as \cite{Fortunati_17}
\begin{equation}
    \lb\bigl(\unk_0\bigr) = \mcrb\bigl(\unk_0\bigr) + \underbrace{\bigl(\bar{\unk} - \unk_0\bigr)\bigl(\bar{\unk} - \unk_0\bigr)^\trans}_{\text{Bias term}},
   \label{eq:lb_def} 
\end{equation}
where $\bar{\unk} = \left[\bar{\pUE}^\trans\ \realPart{\bar{\alphaVect}^\trans}\ \imagPart{\bar{\alphaVect}^\trans}\right]^\trans \in \real{10\times 1}$ denotes the true parameter vector. Since \eqref{eq:lb_def} is conditioned on a residual DPN realization, the expected performance follows by averaging it over the residual DPN distribution in \eqref{eq:pdf_dpnInc} \cite{Richmond_15}. The resulting average position-domain lower bound under model mismatch is given by
$ \lb^{\text{avg}}(\pUE) = \sqrt{\operatorname{tr}([\expecDPNres{\lb(\unk_0)}]_{1: 2, 1: 2})}.$




\section{Simulation Results}\label{sec:results}


In this section, the impact of PN on target position estimation is evaluated. In addition to the proposed \textit{hybrid monostatic-bistatic} architecture, results for a \textit{monostatic-only} scheme are also reported, where only the monostatic observations at the nodes are used.\footnote{A bistatic-only baseline is omitted because, in the considered two-node configuration, both bistatic observations depend on the target position through the same delay, $\tau_{A,B}=\tau_{B,A}$, making the position unidentifiable.} Unless stated otherwise, simulations use the parameter values listed in Table~\ref{tab:sim-config}. The data vectors $\x{A}$ and $\x{B}$ are independently drawn from a QPSK alphabet and satisfy $\norm{\x{A}}^2\!=\!\norm{\x{B}}^2\!=\!\Nsc P_{\text{Tx}}$. The channel gains are given by $\alpha_{p}\! =\! \frac{\lambda}{(4\pi)^{3/2}\,\norm{\p{p} - \pUE}^2}e^{-j[2\pi\, f_c \tau_{p} \!- \!\beta_p]}$ and $\alpha_{p, q} = \frac{\lambda}{(4\pi)^{3/2}\, \norm{\p{p} - \pUE}\norm{\p{q} - \pUE}}e^{-j[2\pi\, f_c \tau_{p, q} \!-\! \beta_{p, q}]}$, where $\beta_p$ and $\beta_{p,q}$ are drawn from a uniform distribution over $[-\pi,\pi)$.



\begin{table}
\centering
\caption{Simulation parameters.}
\label{tab:sim-config}
\begin{tabular}{|c|c|} 
    \hline
    \textbf{Parameter} & \textbf{Value} \\
    \hline
    Carrier frequency ($f_c$) & 28 GHz \\
    \hline
    Total bandwidth ($B$) & 50 MHz\\
    \hline
    Number of subcarriers ($\Nsc$)& 256\\
    \hline
    Subcarrier spacing ($\Delta f$) & 195.31 kHz\\
    \hline
    Symbol duration ($T$) & 5.12 $\mu$s\\
    \hline
    Cyclic prefix duration ($\Tcp$) & 1.28 $\mu$s\\
    \hline
    Total symbol duration ($\Tsym$) & 6.40 $\mu$s\\
    \hline
    Noise power ($\nVar$) & -97 dBm\\
    \hline
    Position of node A ($\p{A}$) & [20, 15] m\\
    \hline
    Position of node B ($\p{B}$) & [10, 40] m\\
    \hline
    Target position ($\pUE$) & [30, 20] m\\
    \hline
    Oscillator 3-dB Bandwidth ($\fosc{A} = \fosc{B}$) & 100 kHz\\
    \hline
    Transmit power ($P_{\text{Tx}}$) & 20 dBm\\
    \hline
\end{tabular}
\end{table}

\subsection{Performance with Respect to Transmit Power}

 Fig.~\ref{fig:res_Ptx} shows the position estimation performance as a function of the transmit power for $\fosc{A} = \fosc{B}=100\ \text{kHz}$. Comparing the CRB and HCRB for both architectures shows that PN only slightly degrades the attainable accuracy when the receiver is aware of its presence. In contrast, neglecting PN introduces a model mismatch that limits the estimation performance, causing the LB to saturate at a moderate transmit power. At low transmit power, the estimation performance is dominated by thermal noise, whereas as transmit power increases, PN is no longer masked by thermal noise and the bias term in \eqref{eq:lb_def}, which is independent of the signal-to-noise ratio (SNR), dominates the LB. The stronger impact observed in the hybrid architecture is due to the bistatic components, which introduce PN-impaired signal terms with larger residual PN variance because they do not benefit from \textit{common-oscillator cancellation}, as discussed in Sec.~\ref{subsec:PN-stat} after \eqref{eq:cov_Delta} (i.e., \textit{delay-dependent} monostatic PN vs. \textit{delay-independent} bistatic PN). Consequently, despite the additional position information provided by the bistatic observations, the PN-ignorant receiver performs better with the monostatic-only scheme, indicating that the larger bistatic PN variance outweighs this information gain. When PN is accounted for, the hybrid architecture outperforms the monostatic-only scheme over the considered transmit-power range, showing that bistatic observations continue to provide useful position information despite the residual PN.
 


\subsection{Performance with Respect to Oscillator Quality}

Fig.~\ref{fig:res_oscBw} shows the impact of oscillator quality on position estimation performance. Decreasing the oscillator quality, i.e., increasing $f_{\text{3dB}}$, significantly degrades the LB due to model mismatch, while the HCRB remains close to the CRB by exploiting PN statistics. At large oscillator bandwidths, the HCRBs of the two architectures tend to saturate, suggesting that the attainable accuracy becomes weakly sensitive to further oscillator degradation. As in Fig.~\ref{fig:res_Ptx}, neglecting PN has a stronger impact on the hybrid scheme than on the monostatic-only scheme, whereas accounting for PN enables the hybrid architecture to outperform its monostatic-only counterpart.

\begin{figure}
    \centering
    \setlength\figurewidth{0.9\linewidth}
    \setlength\figureheight{0.55\linewidth}
%
\definecolor{mycolor1}{rgb}{0.00000,0.44700,0.74100}%
\definecolor{mycolor2}{rgb}{0.85000,0.32500,0.09800}%
\definecolor{mycolor3}{rgb}{0.92900,0.69400,0.12500}%
\definecolor{mycolor4}{rgb}{0.28627,0.85882,0.25098}%
\definecolor{mycolor5}{rgb}{0.81961,0.01569,0.54510}%
\definecolor{mycolor6}{rgb}{0.52157,0.08627,0.81961}%
\definecolor{mycolor7}{rgb}{0.12941,0.12941,0.12941}%
\begin{tikzpicture}

\begin{axis}[%
width=0.951\figurewidth,
height=\figureheight,
at={(0\figurewidth,0\figureheight)},
scale only axis,
xmin=-10,
xmax=55,
xlabel style={font=\scriptsize\color{mycolor7}, yshift=3pt},
xlabel={Transmit Power [dBm]},
ymode=log,
ymin=0.0002,
ymax=1,
yminorticks=true,
ylabel style={font=\scriptsize\color{mycolor7}, yshift=-4pt},
ylabel={PEB [m]},
axis background/.style={fill=white},
xmajorgrids,
ymajorgrids,
yminorgrids,
legend columns=3,
transpose legend,
legend style={
    at={(0.01,0.02)},
    anchor=south west,
    legend cell align=left,
    column sep=0.15em,
    font=\scriptsize,
    legend image post style={xscale=0.5},
    /tikz/every even column/.append style={column sep=0.05em}
},
xticklabel style={
        font=\scriptsize
    },
yticklabel style={
        font=\scriptsize
    },
]
\addplot [color=mycolor1, line width=2.0pt]
  table[row sep=crcr]{%
-20	1.22280998731874\\
-10	0.386686470552882\\
0	0.122280998731874\\
10	0.0386686470552881\\
20	0.0122280998731874\\
30	0.00386686470552883\\
40	0.00122280998731874\\
50	0.000386686470552882\\
60	0.000122280998731874\\
};
\addlegendentry{CRB - Hybrid}

\addplot [color=mycolor3, line width=2.0pt]
  table[row sep=crcr]{%
-20	1.2134168665134\\
-10	0.386421022754742\\
0	0.124765767187286\\
10	0.0412202984632859\\
20	0.0141227612898054\\
30	0.00491929430104556\\
40	0.00165625579827441\\
50	0.000535690814574778\\
60	0.000170368540753777\\
};
\addlegendentry{HCRB - Hybrid}

\addplot [color=mycolor2, line width=2.0pt]
  table[row sep=crcr]{%
-20	2.02414590569791\\
-10	0.662404600978065\\
0	0.270076240759956\\
10	0.190675926565433\\
20	0.180828432451556\\
30	0.179814020397417\\
40	0.179712267463601\\
50	0.179702082649945\\
60	0.179701062065054\\
};
\addlegendentry{LB - Hybrid}
\addplot [color=mycolor4, dashed, line width=2.0pt]
  table[row sep=crcr]{%
-20	2.28730159353315\\
-10	0.723308273129741\\
0	0.228730159353316\\
10	0.0723308273129745\\
20	0.0228730159353314\\
30	0.00723308273129744\\
40	0.00228730159353317\\
50	0.000723308273129743\\
60	0.000228730159353316\\
};
\addlegendentry{CRB - Monostatic}

\addplot [color=mycolor6, dashed, line width=2.0pt]
  table[row sep=crcr]{%
-20	2.28781662929671\\
-10	0.724788485531222\\
0	0.231865970339757\\
10	0.0759042459100261\\
20	0.0255790573299869\\
30	0.00889159581748521\\
40	0.00300329718313241\\
50	0.00097074417978553\\
60	0.000308340115594391\\
};
\addlegendentry{HCRB - Monostatic}

\addplot [color=mycolor5, dashed, line width=2.0pt]
  table[row sep=crcr]{%
-20	2.62538058560765\\
-10	0.831684151531005\\
0	0.267593084309137\\
10	0.0979633189326457\\
20	0.0562076385869921\\
30	0.0501896766302966\\
40	0.0506626751819678\\
50	0.0505706952321523\\
60	0.0505426242374069\\
};
\addlegendentry{LB - Monostatic}
\end{axis}

\begin{axis}[%
width=1.227\figurewidth,
height=1.227\figureheight,
at={(-0.16\figurewidth,-0.135\figureheight)},
scale only axis,
xmin=0,
xmax=1,
ymin=0,
ymax=1,
axis line style={draw=none},
ticks=none,
axis x line*=bottom,
axis y line*=left
]
\end{axis}
\end{tikzpicture}%
    \caption{Performance bounds on position accuracy with respect to transmit power for $\fosc{A} = \fosc{B}=100\ \text{kHz}$.}
    \label{fig:res_Ptx}
\end{figure}
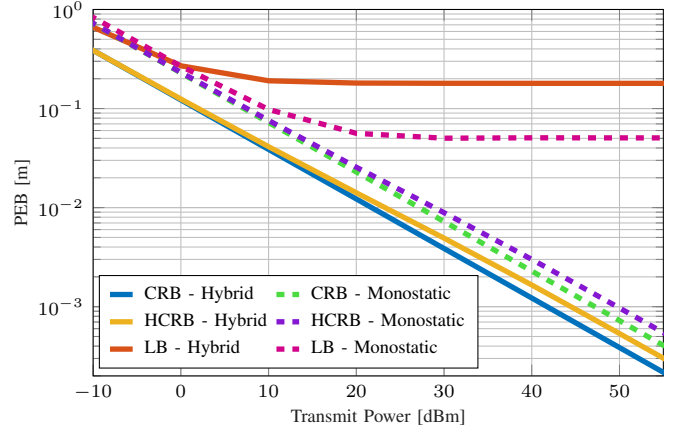
\begin{figure}
    \centering
    \setlength\figurewidth{0.9\linewidth}
    \setlength\figureheight{0.55\linewidth}
%
\definecolor{mycolor1}{rgb}{0.00000,0.44700,0.74100}%
\definecolor{mycolor2}{rgb}{0.85000,0.32500,0.09800}%
\definecolor{mycolor3}{rgb}{0.92900,0.69400,0.12500}%
\definecolor{mycolor4}{rgb}{0.28627,0.85882,0.25098}%
\definecolor{mycolor5}{rgb}{0.81961,0.01569,0.54510}%
\definecolor{mycolor6}{rgb}{0.52157,0.08627,0.81961}%
\definecolor{mycolor7}{rgb}{0.12941,0.12941,0.12941}%
\begin{tikzpicture}

\begin{axis}[%
width=0.951\figurewidth,
height=\figureheight,
at={(0\figurewidth,0\figureheight)},
scale only axis,
xmode=log,
xmin=1000,
xmax=10000000,
xminorticks=true,
xlabel style={font=\scriptsize\color{mycolor7}, yshift=3pt},
xlabel={Oscillator 3-dB Bandwidth [Hz]},
ymode=log,
ymin=0.01,
ymax=20,
yminorticks=true,
ylabel style={font=\scriptsize\color{mycolor7}, yshift=-4pt},
ylabel={PEB [m]},
axis background/.style={fill=white},
xmajorgrids,
xminorgrids,
ymajorgrids,
yminorgrids,
legend columns=3,
transpose legend,
legend style={
    at={(0.01,0.98)},
    anchor=north west,
    legend cell align=left,
    column sep=0.15em,
    font=\scriptsize,
    legend image post style={xscale=0.5},
    /tikz/every even column/.append style={column sep=0.05em}
},
xticklabel style={
        font=\scriptsize
    },
yticklabel style={
        font=\scriptsize
    },
]
\addplot [color=mycolor1, line width=2.0pt]
  table[row sep=crcr]{%
1000	0.0122280998731874\\
10000	0.0122280998731874\\
100000	0.0122280998731874\\
1000000	0.0122280998731874\\
10000000	0.0122280998731874\\
};
\addlegendentry{CRB - Hybrid}

\addplot [color=mycolor3, line width=2.0pt]
  table[row sep=crcr]{%
1000	0.0129230464850412\\
10000	0.0135571953889434\\
100000	0.0141227612898053\\
1000000	0.0153763351929277\\
10000000	0.0162719108330389\\
};
\addlegendentry{HCRB - Hybrid}

\addplot [color=mycolor2, line width=2.0pt]
  table[row sep=crcr]{%
1000	0.0190240800651189\\
10000	0.0559547954881852\\
100000	0.181566339895043\\
1000000	1.23735000865902\\
10000000	21.356835517874\\
};
\addlegendentry{LB - Hybrid}

\addplot [color=mycolor4, dashed, line width=2.0pt]
  table[row sep=crcr]{%
1000	0.0228730159353314\\
10000	0.0228730159353314\\
100000	0.0228730159353314\\
1000000	0.0228730159353314\\
10000000	0.0228730159353314\\
};
\addlegendentry{CRB - Monostatic}

\addplot [color=mycolor6, dashed, line width=2.0pt]
  table[row sep=crcr]{%
1000	0.0231865970339757\\
10000	0.0240030301153204\\
100000	0.0255790573299869\\
1000000	0.0281176948168802\\
10000000	0.0300329718313241\\
};
\addlegendentry{HCRB - Monostatic}

\addplot [color=mycolor5, dashed, line width=2.0pt]
  table[row sep=crcr]{%
1000	0.0234313827012308\\
10000	0.0279070746931326\\
100000	0.0595248310089577\\
1000000	0.580541846059036\\
10000000	7.76045221472338\\
};

\addlegendentry{LB - Monostatic}
\end{axis}

\begin{axis}[%
width=1.227\figurewidth,
height=1.227\figureheight,
at={(-0.16\figurewidth,-0.135\figureheight)},
scale only axis,
xmin=0,
xmax=1,
ymin=0,
ymax=1,
axis line style={draw=none},
ticks=none,
axis x line*=bottom,
axis y line*=left
]
\end{axis}
\end{tikzpicture}%
    \caption{Performance bounds on position accuracy with respect to oscillator bandwidth ($\fosc{A} = \fosc{B}$) for a transmit power of 20 dBm.}
    \label{fig:res_oscBw}
\end{figure}
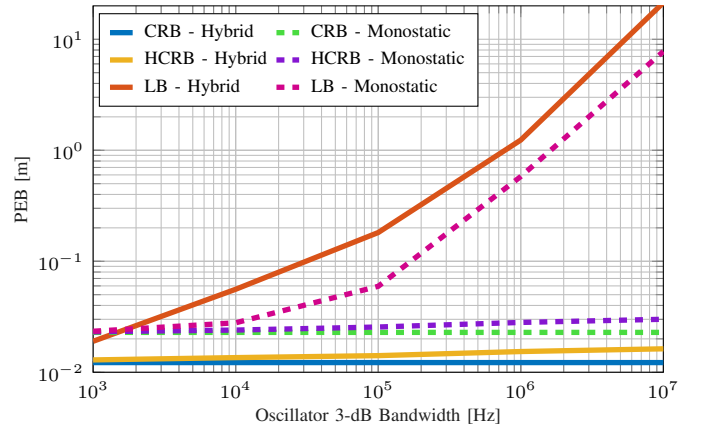

\subsection{Performance with Respect to Target Position}

\begin{figure}[!t]
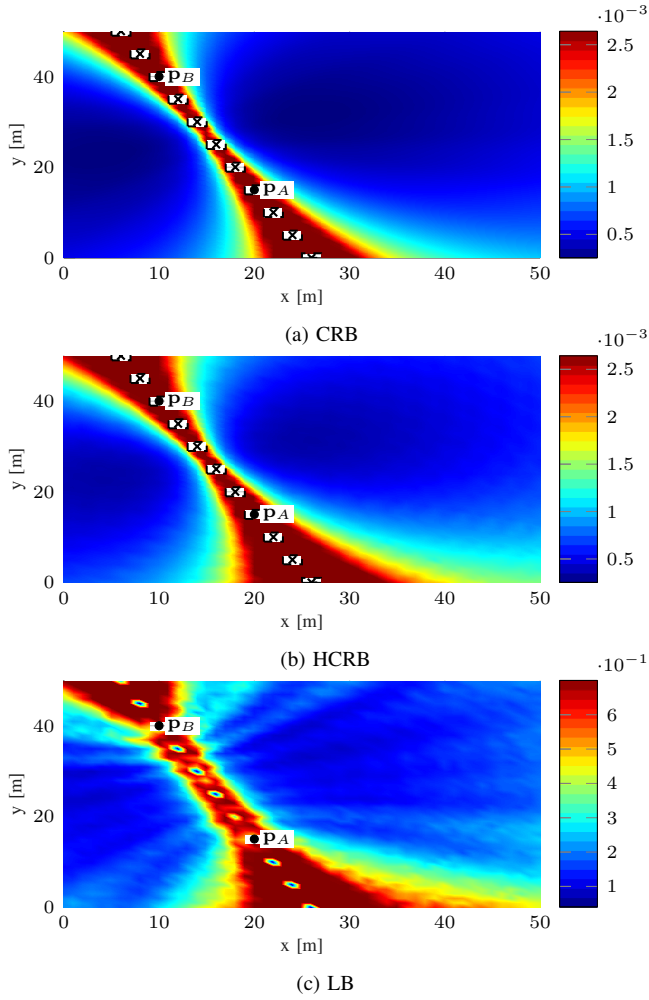

    \centering

    \subfloat[CRB \label{fig:res_pos_crb}]{%
    \begingroup
        \setlength\figurewidth{0.715\linewidth}%
        \setlength\figureheight{0.34\linewidth}%
        \input{figures/hybrid_vs_userPosition_SNR_40_dB_100_kHz_CRB_v2.tex}
    \endgroup
    }
    
    \vspace{-0.6cm}

    \subfloat[HCRB\label{fig:res_pos_hcrb}]{%
    \begingroup
        \setlength\figurewidth{0.715\linewidth}%
        \setlength\figureheight{0.34\linewidth}%
        \input{figures/hybrid_vs_userPosition_SNR_40_dB_100_kHz_HCRB_v2.tex}
    \endgroup
    }

    \vspace{-0.6cm}

    \subfloat[LB\label{fig:res_pos_lb}]{%
    \begingroup
        \setlength\figurewidth{0.715\linewidth}%
        \setlength\figureheight{0.34\linewidth}%
        \input{figures/hybrid_vs_userPosition_SNR_40_dB_100_kHz_MCRB_v2.tex}
    \endgroup
    }
    
    \vspace{0.15cm}

    \caption{Performance bounds on position accuracy with respect to target position for $\fosc{A} = \fosc{B} = 100\ \text{kHz}$ and an SNR of 40 dB.}
    \label{fig:res_pos}
\end{figure}

Fig.~\ref{fig:res_pos} shows the position estimation performance of the hybrid architecture as a function of the 2D target position. In this experiment, the SNR, defined as $\text{SNR} = \norm{\alphaVect}^2/2\Nsc\nVar$, is fixed across the grid rather than the transmit power, so that the spatial variations correspond to the sensing geometry and the location-dependent PN statistics, rather than path loss. The cross markers indicate target positions on the line joining the two sensing nodes, where the position FIM becomes singular because the monostatic delay-gradient vectors, i.e., ${\partial \tau_A}/{\partial \pUE}$ and ${\partial \tau_B}/{\partial \pUE}$, are collinear.

The three maps exhibit similar spatial trends, with regions of low and high PEB occurring approximately at the same target locations for the CRB, HCRB and LB. This suggests that the geometry is the dominant factor determining the position-dependent accuracy. Nevertheless, PN statistics are also location-dependent because the residual DPN covariance depends on the propagation delays, as discussed in Sec.~\ref{subsec:PN-stat}. The HCRB remains close to the CRB by exploiting PN statistics, whereas the LB shows stronger degradation due to the model mismatch. This degradation is slightly more pronounced in regions farther from the nodes, consistent with the delay-dependent monostatic residual PN variance.


\section{Conclusion}


In this work, we investigated the impact of PN on position estimation accuracy in a centralized hybrid monostatic-bistatic OFDM sensing system. The analysis reveals that the common-oscillator cancellation causes the monostatic residual DPN variance to saturate after the round-trip delay, whereas the residual DPN variance of the bistatic links grows linearly with the within-symbol time index. In terms of localization performance, PN-ignorant hybrid processing suffers a stronger localization loss than monostatic-only processing. In contrast, accounting for PN keeps the localization bounds of both architectures close to their PN-free counterparts. Furthermore, in the considered setting, the additional bistatic observations can be exploited to improve localization accuracy relative to monostatic-only sensing at low and medium transmit powers.

As future work, we plan to extend the present framework to multi-target and multi-antenna scenarios, as well as to incorporate clutter effects.

\balance
\bibliographystyle{IEEEtran}
\bibliography{refs}

\end{document}